\documentclass[12pt]{iopart}
\pdfoutput=1
\usepackage{epsfig}
\usepackage{float}
\usepackage{color}
\usepackage{array}
\usepackage{lipsum}
\usepackage{epstopdf}

\begin{document}

\title
[Freezing of ice under an electric field: a simulation study]
{Phase boundaries, nucleation rates and speed of crystal growth of the 
water-to-ice transition under an electric field: a simulation study}

\author{ Alberto Zaragoza }
\address
{Departamento de Estructura de la Materia, Fisica Termica y Electronica,
Facultad de Ciencias Fisicas, Universidad Complutense de Madrid,
28040 Madrid, Spain.}
\address
{Departamento de Ingenieria Fisica, Division de Ciencias e Ingenierias, Universidad de Guanajuato, Loma del Bosque 103, Col. Lomas del Campestre, CP 37150 Leon, Mexico}
\author{ Jorge R.Espinosa }
\address
{Departamento de Qu\'{\i}mica F\'{\i}sica,
Facultad de Ciencias Qu\'{\i}micas, Universidad Complutense de Madrid,
28040 Madrid, Spain.}
\author{ Regina Ramos }
\address
{Centro de Electr\'{o}nica Industrial, Universidad Polit\'{ecnica} de Madrid, 28006 Madrid, Spain}
\author{ Jos\'{e} Antonio Cobos }
\address
{Centro de Electr\'{o}nica Industrial, Universidad Polit\'{ecnica} de Madrid, 28006 Madrid, Spain}
\author{Juan Luis Aragones}
\address
{Condensed Matter Physics Center, Universidad Aut\'{o}noma de Madrid, 28049, Spain}
\author{ Carlos Vega }
\address
{Departamento de Qu\'{\i}mica F\'{\i}sica,
Facultad de Ciencias Qu\'{\i}micas, Universidad Complutense de Madrid,
28040 Madrid, Spain.}
\author{ Eduardo Sanz }
\address
{Departamento de Qu\'{\i}mica F\'{\i}sica,
Facultad de Ciencias Qu\'{\i}micas, Universidad Complutense de Madrid,
28040 Madrid, Spain.}
\ead{esa01@quim.ucm.es}
\author{Jorge Ram\'{i}rez }
\address{Departamento de Ingenier\'{i}a Qu\'{\i}mica Industrial y Medio Ambiente,
Universidad Polit\'{e}cnica de Madrid,
28006 Madrid, Spain}
\author{Chantal Valeriani}\footnote{The last two authors contributed equally to the article}
\address
{Departamento de Estructura de la Materia, Fisica Termica y Electronica, GISC, 
Facultad de Ciencias Fisicas, Universidad Complutense de Madrid,
28040 Madrid, Spain.}


\begin{abstract}

We investigate with computer simulations the effect of applying an electric field
on the water-to-ice transition. We use a combination of state-of-the-art simulation
techniques to obtain phase boundaries and crystal growth rates (direct coexistence), nucleation
rates (seeding) and interfacial free energies (seeding and mold integration).  
First, we consider ice Ih, the most stable polymorph in the absence of a field. 
Its normal melting temperature, speed of crystal 
growth  and nucleation rate (for a given supercooling)
diminish as the intensity of the field goes up. Then, we study 
polarised cubic ice, or ice Icf,  the most stable solid phase under a strong electric field. 
Its normal melting point goes up with the field and, for 
a given supercooling, under the studied field (0.3 V/nm) ice Icf   
nucleates and grows at a similar rate as Ih with  no 
field.  The net effect of the field would be then that ice nucleates at warmer
temperatures, but in the form of ice Icf. The main conclusion of this work is that reasonable 
electric fields (not strong enough to break water molecules apart) are not relevant in the context of homogeneous ice nucleation at 1 bar.

\end{abstract}
\pacs{64.70.-p,61.20.Ja}
\noindent{\it Keywords\/}: electric field, water, phase transitions, nucleation rate \\

\submitto{\JPCM}
\maketitle

\section{Introduction} 

The effect of an applied   electric field on  water's thermal stability and phase transitions  is  still  nowadays a matter of debate \cite{English_2003,English_2006,English_2015}, 
 posing questions such as whether the external field could induce the   appearance of new ice phases,  
 how the ice-water melting temperature could be affected, or whether  the electric field could alter   ice nucleation  or  crystal growth processes.
Answering to the above mentioned questions could improve our understanding  of the microscopic aspects of 
 both  thermodynamics and kinetics of water-ice phase transitions  
under strong electric fields, with potential industrial (macroscopic) applications such as  preventing  
shortcuts on high-voltage power lines \cite{Laforte_1998}, 
food processing, and cryopreservation of cells\cite{Sakai_1990} and living tissues\cite{Wowk_2012, Karlsson_1996}. 


On the one side, previously published experimental results suggested that electric fields  
enhance  self-diffusion of water in confined environments \cite{Diallo_2012} and raise the supercooling 
\cite{Wei_2008, Orlowska_2009} thus affecting both ice nucleation and growth, either promoting \cite{Choi_2005}  or hindering it, depending 
on the charge of the confining surface \cite{Ehre_2010}. 
On the other side, most numerical simulations have investigated 
the effect of very large  external fields (from 5 to 20 V/nm) 
showing that, while promoting  crystallization of polarized ice Ic
\cite{svishchev94,Svishchev_1996,Svishchev_1996a}, it slows down the self-diffusion coefficient,  
  introducing structural changes in liquid water \cite{Kiselev_1996,Jung_1999,Wei_2005,Hu_2011,Hu_2012}.
  
  In Ref.\cite{Aragones_2011a}, the authors numerically studied the effect 
   of   moderately large fields $E$ (0.15-0.3 V/nm, still larger than the dielectric 
strength of real water, $0.06-0.07$ V/nm)\cite{CRC92} on the ice-water phase 
diagram of the TIP4P/2005 water model\cite{JCP_2005_123_234505},  concluding that the main effect of the field was to 
displace the ice-water phase boundaries, increasing the thermodynamic stability of phases with higher dielectric constants. 
In particular, they predicted that a field of 0.3 V/nm shifted the melting point of ice Ih towards lower values, at low 
pressures. In the same work, the authors suggested that cubic ice (Ic) could become more stable 
than ice Ih for  a field of 0.15 V/nm at 1 bar, given that the structure of ice 
Ic would allow the full saturation of the polarisation ($\langle M\rangle/(N \ \mu_{eff}) = 1$, 
where $\mu_{eff}$ is the effective dipole moment of the model, $N$ the total number of ice molecules and $\langle M\rangle$ the average polarisation). 
Later on, Yan and Patey \cite{Yan_2011,Yan_2012} studied heterogeneous ice nucleation of six-site\cite{nada03} and TIP4P/ICE\cite{JCP_2005_122_234511} water models
under moderately large electric fields with magnitudes up to 2.5 V/nm, applied  within a narrow slab-like region 
(10-20 \AA). They showed that these fields speeded up  ice nucleation 
in the proximity of that region, and reported  not only ice nucleation occurring for  TIP4P/ICE  at temperatures 
as high as the melting temperature, \emph{i.e.} $T_m=270$ K, but also the growth of a dipole disordered cubic phase (Ic) 
away from that region. In a later work, the same authors \cite{Yan_2014} showed that uniform electric fields on the
order of 1-2 V/nm increase the melting point of the six-site water model\cite{nada03} and concluded that polarized water can be deeply 
supercooled under an electric field without cooling to very low temperatures and thus freeze on simulation scales. They also suggested that the field might 
reduce the surface tension of the ice/liquid interface, but didn't provide any evidence.

The aim of our work is to give  a comprehensive and conclusive study on the effect of a constant electric field on the ice-water coexistence
temperature at ambient pressure, on ice nucleation  and crystal growth  of two ice polymorphs: hexagonal ice Ih, 
the most stable polymorph in the absence of field at ambient pressure, 
and cubic Icf (the ferroelectric version of cubic ice Ic), the most stable polymorph when a large
electric field is applied.

\section{Simulation details}


Throughout our study, we  simulate water via TIP4P/ICE\cite{JCP_2005_122_234511} by means of 
the GROMACS Molecular Dynamics package \cite{JMM_2001_07_0306} in the NpT ensemble, where the pressure 
is fixed at 1 bar. 
In order to fix  pressure and  temperature, we make use of the anisotropic Parrinello-Rahman barostat 
\cite{JAP_1981_52_007182} and  Nose-Hoover thermostat \cite{MP_1984_52_0255,PRA_1985_31_001695}, 
respectively, both with a relaxation time of 1 ps. 
A leap-frog algorithm\cite{Hockney_1981} is used for the integration of the equations of motion, with a time step of 2 fs. 
Electrostatic interactions are calculated by means of particle-mesh Ewald\cite{JCP_1995_103_8577}. The real part of the 
electrostatic potential and the Lennard-Jones interaction are cut-off at 9 \AA, and long tail corrections are added to the Lennard-Jones interactions.

\begin{figure}[h!]\centering
\includegraphics[clip,width=1.0\textwidth,angle=-0]{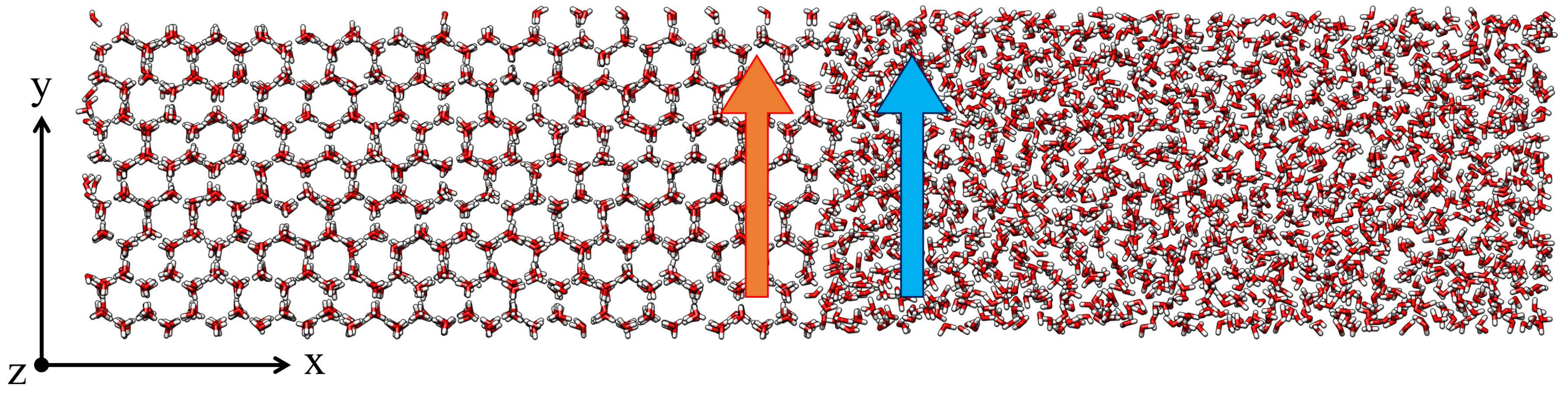}
\caption{Snapshot of the direct coexistence simulation box of ice Ih. The direction of growth ($x$) is perpendicular to the secondary prismatic plane. 
The external electric field is applied in the direction $y$, parallel to the interface (blue arrow). When the ice phase is polarized, 
the corresponding permanent polarisation vector (orange arrow) is parallel to the external field.}
\label{Fig_snapshot}
\end{figure}

To obtain the melting temperature and the growth rate of ice, both in the absence and presence of an external field, we use the direct 
coexistence method \cite{ladd77,ladd78,Garc_a_Fern_ndez_2006}. In the method, a solid phase is brought into contact with a liquid phase 
at several temperatures. The melting point is determined as the highest temperature at which the crystal slab does not melt. 
In Figure \ref{Fig_snapshot} a typical 
snapshot of the simulation box during a direct coexistence simulation is shown, along with the typical orientations of the electric 
field (in blue) and the polarisation (in orange) vectors used in this work.

In order to compute the nucleation rate we  employ the Seeding technique \cite{Espinosa_2014b,Espinosa_2016,Espinosa_2016b,Sanz_2013} that combines 
Classical Nucleation Theory (CNT)\cite{kelton,ZPC_1926_119_277_nolotengo,becker-doring} with numerical calculations. 
Having equilibrated    a spherical ice cluster of a given size ($N_c$) embedded in supercooled water, 
we follow  the  time evolution of the cluster size at different temperatures to  estimate the   temperature $T_c$ at which 
the cluster is critical   (i.e. the temperature enclosed between the highest one at which the cluster grows and the lowest at which it melts). Care must be taken in tuning  
the order
parameter used to detect the crystal cluster size (i.e. number of ice molecules)\cite{Zaragoza_2015,Lechner_2008}. Details are provided in the Supporting Information.

According to the Classical Nucleation Theory, the critical cluster size is expressed as 
\begin{equation}  
\label{ennec}
 N_c = \frac{32 \pi \gamma^3}{3 \rho_s^2 |\Delta \mu|^3}
\end{equation}
where $\rho_s$   is the ice density, $\Delta \mu$ the chemical potential difference between 
ice and water, and $\gamma$ the ice-water interfacial free-energy. 
Having computed $N_c$, we can evaluate the interfacial free-energy $\gamma$ of a spherical cluster of a given size (i.e. at a temperature below
coexistence) via Eq. \ref{ennec} by computing  $\rho_s$ 
(via NpT simulations) and  $\Delta \mu = \mu_{ice} - \mu_{water}$ (via thermodynamic integration of the Gibbs-Helmholtz equation from the melting temperature to $T_c$)\cite{Frenkel_2001}. Note that in the presence of an external field, the term of the Hamiltonian corresponding to the interaction between the field and the system polarisation 
has not been considered when integrating the enthalpy difference between the solid and the liquid to obtain $\Delta \mu$.

Knowing the number density of critical clusters, $\rho_f \exp\left(-\Delta G_c/k_B T\right)$
(where $\rho_f$ is the liquid density and  $\Delta G_c = \frac{N_c |\Delta \mu|}{2}$ the nucleation free-energy barrier height), 
the CNT expression for the nucleation rate ($J$) is 
 \begin{equation}  
\label{rate}
J = \sqrt{\frac{|\Delta \mu|}{6 \pi k_b T N_c}} f^{+} \rho_f \exp{(-\Delta G_c/k_B T)}
\end{equation}
obtained by multiplying the number density of critical clusters by a kinetic prefactor  $\sqrt{\frac{|\Delta \mu|}{6 \pi k_b T N_c}} f^{+}$, 
where\cite{Kelton_2010,Espinosa_2014b}
  \begin{equation}  
\label{fmas}
f^{+}(T) = \frac{24 D(T) N_c(T)^{2/3}}{\lambda^2}, 
\end{equation}
is the the attachment rate of particles to the critical cluster, 
$D$ the liquid diffusion coefficient and $\lambda$ the distance travelled by a particle to attach to the cluster's surface
($\lambda$ is typically one molecular diameter; here, we use $\lambda=4$ \AA \ as in previous work\cite{Espinosa_2014b}).

For the calculation of the ice-water interfacial free-energy at coexistence, we use the Mold Integration method \cite{Espinosa_2014},  
based on computing the reversible work $\Delta G$ needed to induce the formation of a crystal slab embedded in liquid water, 
related to the interfacial free-energy at coexistence by $\Delta G=2A\gamma$, 
where $2A$ corresponds to the area of the two crystalline interfaces of the mold. The formation of the crystalline slab 
is induced by switching on an attractive interaction between the fluid particles and the mold's potential energy wells, 
located at the equilibrium positions of the oxygen atoms of the ice lattice plane of interest\cite{Espinosa_2014,Espinosa_2016c}.

In all cases dealing with the ice phase with a permanent polarisation (Icf), the electric field 
was applied in the direction parallel to the polarisation vector. 
When computing the melting temperature via direct coexistence simulations, we exposed the secondary prismatic plane to the 
liquid water and oriented the electric field parallel to the interface (see Figure \ref{Fig_snapshot}).
When computing the nucleation rate via the Seeding method the relative orientation of the electric field with respect to the interface is irrelevant, 
due to the spherical symmetry of the crystalline cluster.


\section{Results}

\subsection{Ice Ih-water phase diagram and nucleation of ice Ih}

Inspired by previous works\cite{Aragones_2011a,Yan_2011,Yan_2012,Yan_2014}, we first compute the melting temperature of ice Ih, the most stable ice polymorph at ambient pressure, under a constant  
electric field by means of direct coexistence simulations\cite{ladd77,ladd78}. 
\begin{figure}[h!]\centering
\includegraphics[clip,width=0.7\textwidth,angle=-0]{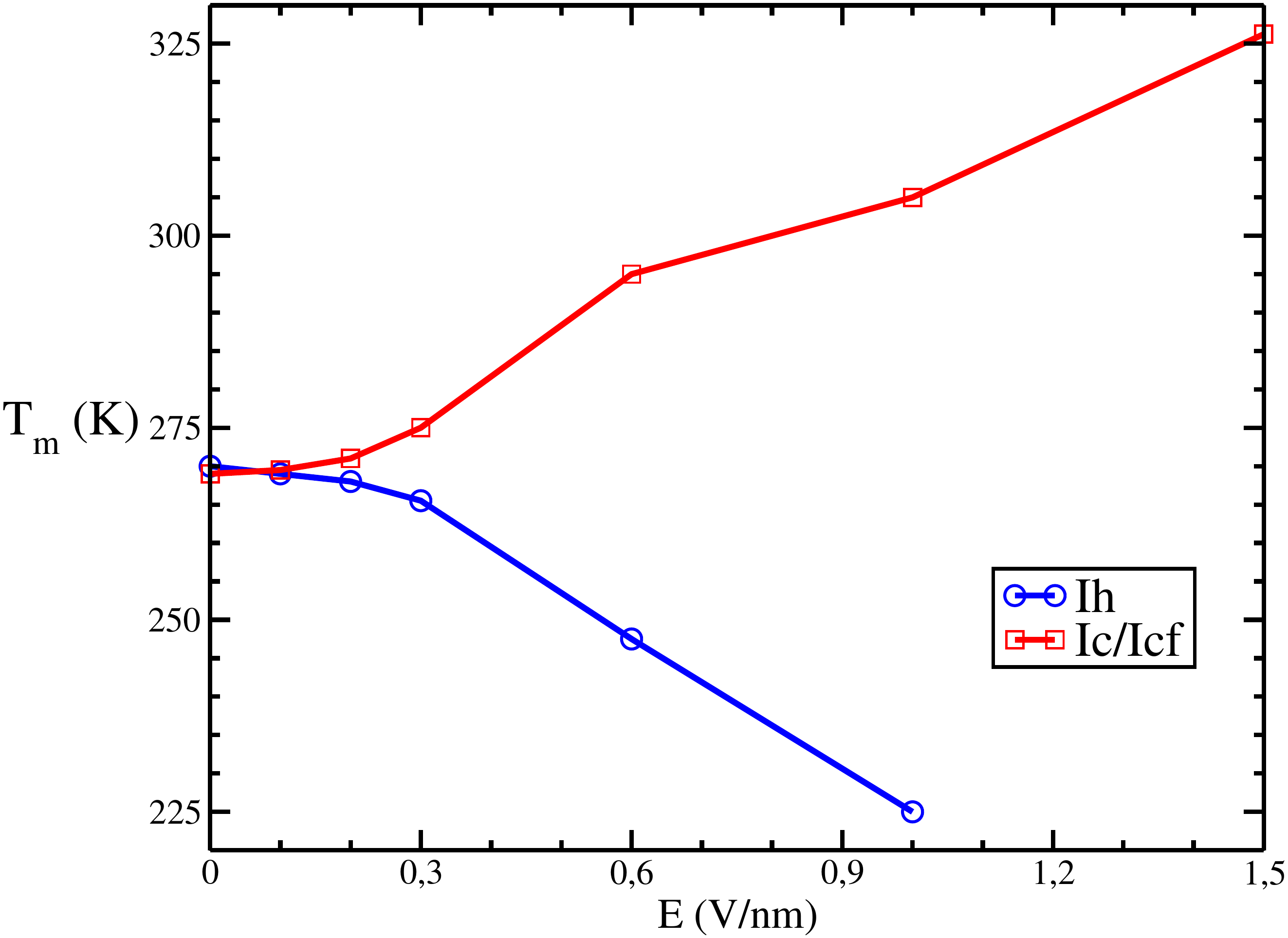}
\caption{Melting temperatures ($T_m$) of the Ih (in blue) and Ic/Icf (in red) ice phases at 1 bar as a function of the magnitude of the 
applied electric field $E$.}
\label{Fig_phase}
\end{figure}

As shown in Fig. \ref{Fig_phase} (blue line),  when $E<0.15$ V/nm the ice Ih melting temperature is only slightly affected by the 
presence of the field (given that  the thermal energy is still high with respect to the energy of the molecules under the applied  field).  
As soon as  $E>0.15$ V/nm, $T_m$ decreases down until $E=1.0$ V/nm.
For fields larger than 1.0 V/nm, the temperature of coexistence drops down so much that establishing the melting temperature by means of direct coexistence becomes too expensive numerically. 
Therefore, the electric field decreases the melting temperature, thus hindering freezing of ice Ih.
This decay of $T_m$ is expected, since the dielectric constant of liquid water is 
slightly higher than that of ice Ih in the TIP4P/ICE model\cite{Aragones_2011}, and therefore 
water is stabilised under a large electric field. Note that this is not the case in real experimental water,
where the dielectric constant of ice Ih is slightly higher than that of liquid water.

In order to understand how an applied electric field  affects the nucleation of ice Ih 
from supercooled water we compute the nucleation rate 
under an applied field of  0.3 V/nm,  corresponding to the smallest value 
at which the effect of the field on $T_m$ is clearly detectable, and compare it to the rate of ice Ih without any applied field. 
To compute the nucleation rate of the desired ice polymorph we make use of the Seeding technique\cite{Espinosa_2014b,Espinosa_2016,Espinosa_2016b,Sanz_2013}. 

Having established the melting temperature of ice Ih for 0 V/nm (270$\pm$1 K) and 0.3 V/nm 
(265.5$\pm$1 K, see Figure \ref{Fig_phase}), 
we first compute the chemical potential difference between the supercooled liquid and ice Ih using thermodynamic 
integration\cite{Frenkel_2001}.
\begin{figure*}[!htb]\centering
\includegraphics[width=0.7\textwidth,clip=]{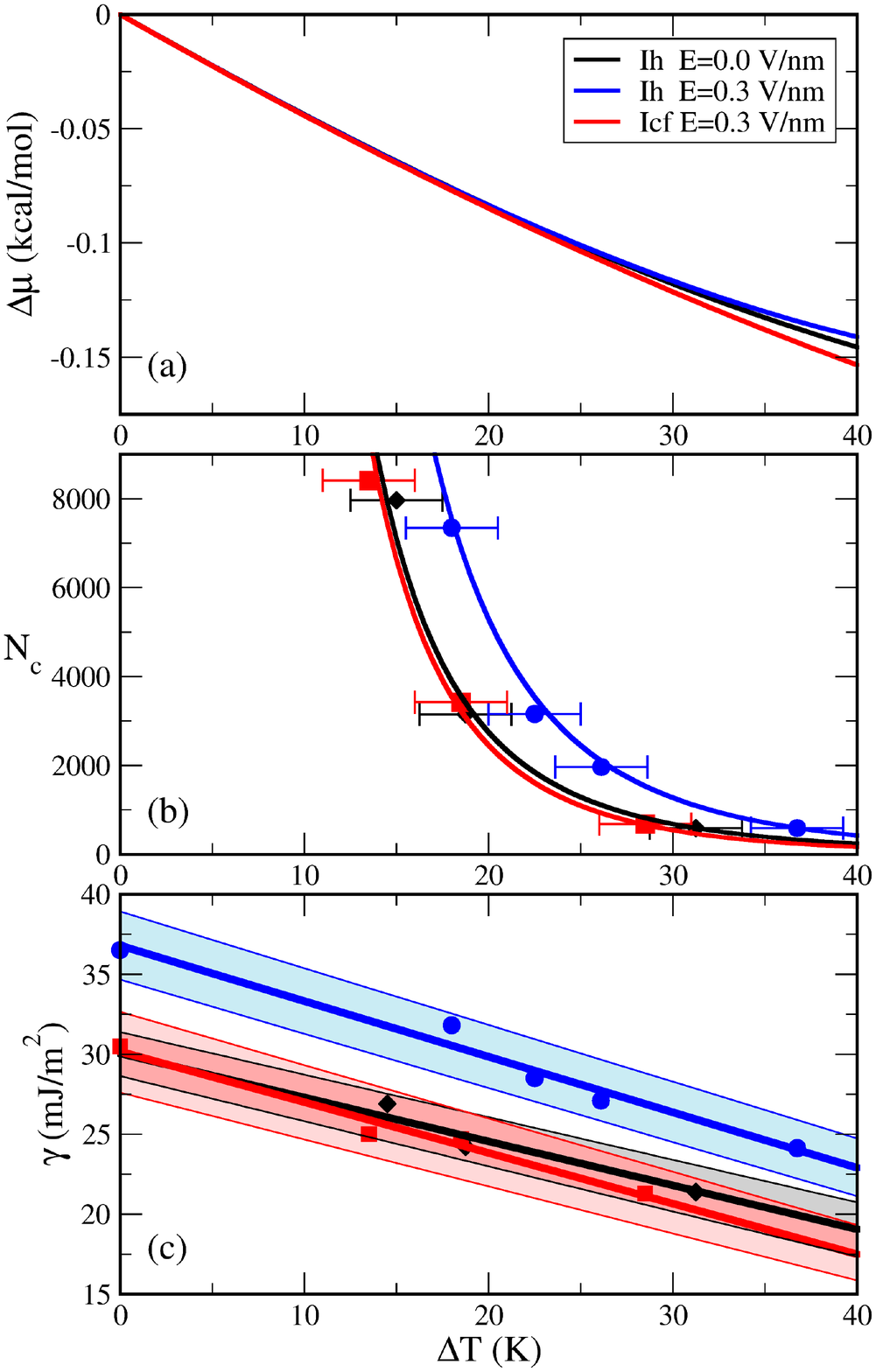}
\caption{a) Chemical potential difference ($\Delta \mu$) between the liquid and ice Ih without the field (in black), 
ice Ih with the field of 0.3 V/nm (in blue) and ice Icf with the same field (in red), as a function of supercooling 
$\Delta T= T_m - T$. $N_c$ (b) and 
ice-water interfacial free-energy (c)  as a function of  supercooling for the different ice polymorphs and electric fields 
(see legend). The same  color code applies to all graphs. Color bands indicate error bars.}
\label{Anuclea}
\end{figure*}
As shown in Figure \ref{Anuclea} (a), the values of $\Delta \mu$ without (in black) and with the field (in blue) are very similar down to a supercooling of about 30 K. 
Therefore, for a given supercooling, the thermodynamic driving force for nucleation of ice Ih is not affected by the presence of the electric field. 

Next, we prepare an initial configuration for the Seeding calculations, 
as described in Ref. \cite{Espinosa_2015},
and establish 
 the temperature at which each embedded ice cluster is critical. 
Care must be taken when preparing the  configuration in the presence of the electric field, given that the surrounding 
water must be  allowed to polarize and equilibrate properly. 
Figure \ref{Anuclea} (b) shows the critical cluster sizes $N_c$ as a function of supercooling. 
For a given size, clusters of ice Ih under 0.3 V/nm (in blue) are critical at a larger supercooling than 
ice Ih clusters without the field (in black). 

Making use of Eq.\ref{ennec}, knowing $\Delta \mu$, $N_c$ and the density of the solid phase
we now compute the ice-water interfacial free-energy (Figure \ref{Anuclea} (c) and Table \ref{table0}). 
At every supercooling,  $\gamma_{Ih}(E=0.0 $ V/nm) is lower than $\gamma_{Ih}(E=0.3$ V/nm).  
The same applies  at coexistence, where $\gamma$ is computed via the Mold Integration technique \cite{Espinosa_2014}, 
averaging not only over the three crystallographic planes (prismatic, secondary prismatic, basal) but also over the three relative orientations of the applied field with respect 
to the direction perpendicular to the plane. 
 
\begin{table}[h!]
\centering
\begin{tabular}{ccccccccccc}
Ice & $N$ & $N_c$ & $\Delta T$ & $\Delta \mu$ & $\rho_f$ & $\gamma$ & $\Delta G_c$ & $D$ & $f^+$ & $\log(J)$ \\
\hline
Ih  &  22712 &  588 & 36.75 &  0.137 & 0.954 & 24.1 &  89 & $0.1 \cdot10^{-10}$ & $2.2\cdot10^{11}$ & -1   \\
Ih  &  76845 & 1964 & 26.1  &  0.103 & 0.966 & 27.1 & 213 & $0.39\cdot10^{-10}$ & $1.7\cdot10^{12}$ & -54  \\
Ih  &  76781 & 3160 & 22.5  &  0.093 & 0.970 & 28.5 & 302 & $0.57\cdot10^{-10}$ & $3.6\cdot10^{12}$ & -93  \\
Ih  & 182585 & 7348 & 18.0  &  0.076 & 0.975 & 31.8 & 611 & $0.89\cdot10^{-10}$ & $6.0\cdot10^{12}$ & -227 \\
\hline
\hline
Icf &  17709 &  680 & 28.5  &  0.115 & 0.975 & 21.1 &  80 & $0.89\cdot10^{-10}$ & $0.1\cdot10^{13}$ & 3   \\
Icf &  63178 & 3420 & 18.5  &  0.078 & 0.986 & 24.7 & 262 & $0.2 \cdot10^{-9}$  & $0.7\cdot10^{13}$ & -75  \\
Icf & 123417 & 8410 & 13.5  &  0.058 & 0.991 & 25.0 & 469 & $0.29\cdot10^{-9}$  & $1.9\cdot10^{13}$ & -165 \\
\end{tabular}
\caption{System size $N$, critical cluster size $N_c$, supercooling $\Delta T$ (in K), chemical potential difference between ice and water $\Delta \mu$ (in kcal/mol), fluid density $\rho_f$ (in $g/cm^3$), liquid-ice interfacial free-energy $\gamma$ (in mJ/$m^2$), free-energy barrier height $\Delta G_c$ (in units of $k_BT$), diffusion coefficient $D$ (in $m^2s^{-1}$), attachment rate $f^{+}$ (in $s^{-1}$) and decimal logarithm of the nucleation rate $J$ (in $m^{-3} s^{-1}$) for ice Ih at 0.3 V/nm (top) and ice Icf at 0.3 V/nm (bottom).}
\label{table0}
\end{table}
 
\begin{figure*}[!htb]
\includegraphics[width=0.7\textwidth,clip=]{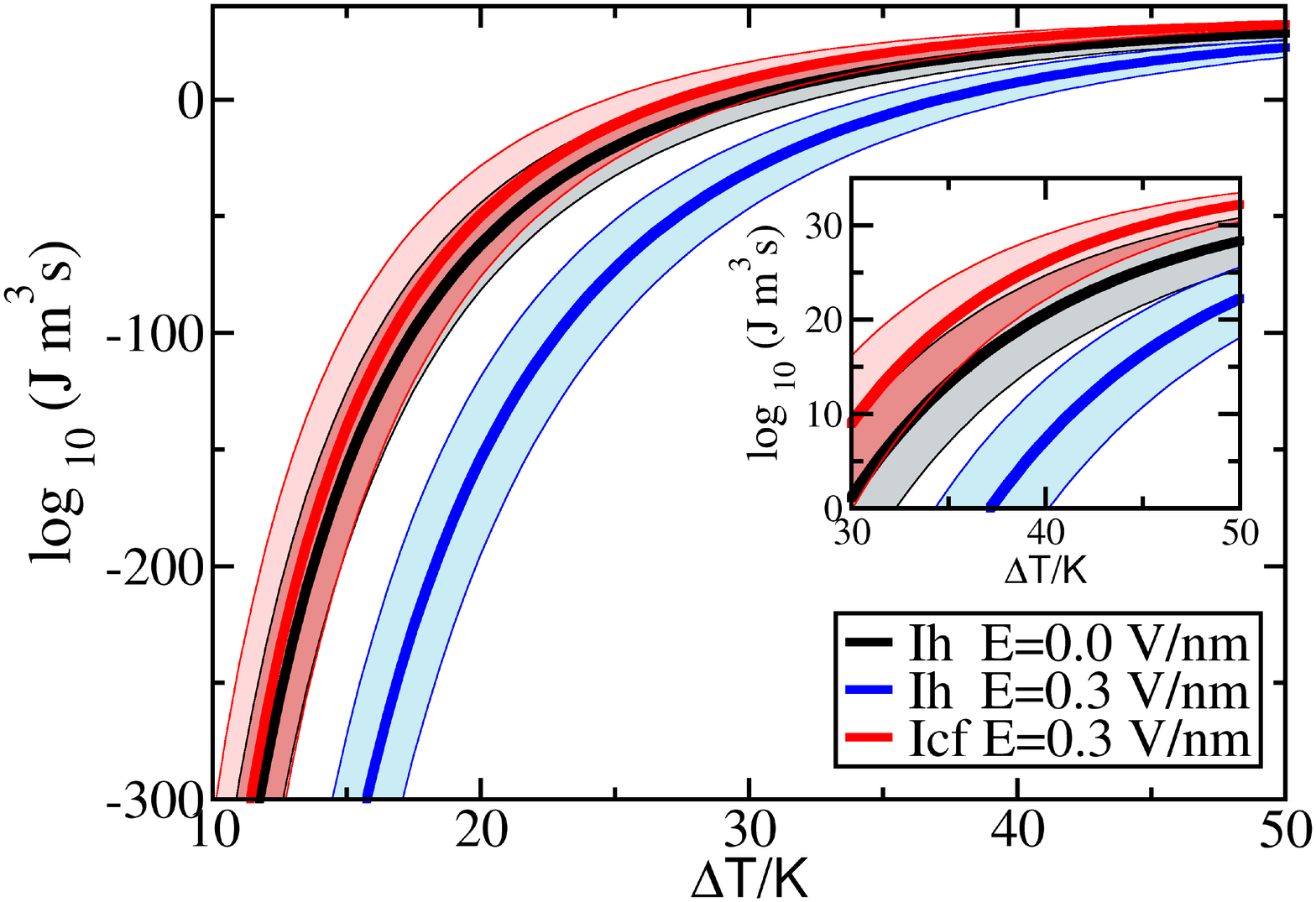}
\caption{Nucleation rate $J$ as a function of  supercooling $\Delta T= (T_m-T)$ for ice Ih  without applying any  field (in black),  
applying an electric field of 0.3 V/nm (in blue)  and ice Icf  applying the same electric field (in red). Color bands are a measure 
of the error bars considering only the statistical uncertainty.}
\label{Bnuclea}
\end{figure*}

Having estimated the free-energy barrier, we can calculate the attachment rate $f^+$ as\cite{Auer_2004}:
\begin{equation}
\label{diffplus}
f^+=\frac{\langle \left(N(t)-N_c\right)^2\rangle}{2t}
\end{equation}
Equation \ref{diffplus} has been used to determine the attachment rate for the first case shown on Table \ref{table0}. 
Then, we use Equation \ref{fmas} for the same case in order to estimate $\lambda\approx 4 \AA$. Fixing the value of $\lambda$, 
and using the values of $D$ and $N_c$ obtained in our simulations
($N_c(T)$ in Eq. \ref{fmas} is obtained from a linear fit to $\gamma(T)$ and Eq. \ref{ennec}), 
we calculate $f^+$ using Equation \ref{fmas} (Table \ref{table0}).
Results of the above mentioned quantities for ice Ih without any field have been already reported in Refs. \cite{Espinosa_2016b,Espinosa_2016}.
To conclude, we  compute the nucleation rate by means of Eq. \ref{rate} (Table \ref{table0} and Figure \ref{Bnuclea}).

For the same supercooling, the nucleation rate of ice Ih with an external field of 0.3 V/nm (in blue) is always lower 
than that without the field (in black).
Therefore the electric field hinders nucleation of ice Ih, given that, in the presence of the field, a deeper 
supercooling is needed to get the same nucleation rate. Given that the values of $\Delta \mu$ and $f^+$ (or $D$) are similar
to those in the absence of a field, 
the nucleation of ice Ih slows down due to the increase in the water-ice Ih interfacial free-energy $\gamma$ when an electric field is applied.

The raise of $\gamma$ can be understood by comparing the orientation imparted by the field on the water molecules in the liquid with 
that of the molecules in the ice Ih crystal. In ice Ih, the total polarisation is zero and the orientation of molecules is 
essentially random. In the absence of field, the molecules in the liquid also have a random orientation, not entirely equal 
to the crystal but structurally similar. However, when a strong field is applied, on the time scale of the simulation the 
ice Ih crystal remains unpolarized, whereas the water molecules in the liquid strongly align their dipoles in the direction 
of the field. Therefore, the orientational distribution of dipoles of ice Ih and liquid water becomes very different. 
Thus, we hypothesize that
the structural difference between ice Ih and polarized water gives rise to a 
sharp increase in the value of the interfacial free-energy ($\gamma$, see Figure \ref{Anuclea} (c)) that hinders the nucleation 
rate ($J$, see Figure \ref{Bnuclea})

Considering the effects of the applied field on both the melting point, which drops with increasing magnitude of the field, 
and the nucleation rate, which for a field $E=0.3$ V/nm and the same supercooling also drops by several orders of magnitude, 
our simulations show that ice Ih freezing is strongly impeded by the field.

\subsection{Ice Icf-water phase diagram and nucleation of ice Icf}

\begin{figure*}[!htb]
\includegraphics[width=1.0\textwidth,clip=]{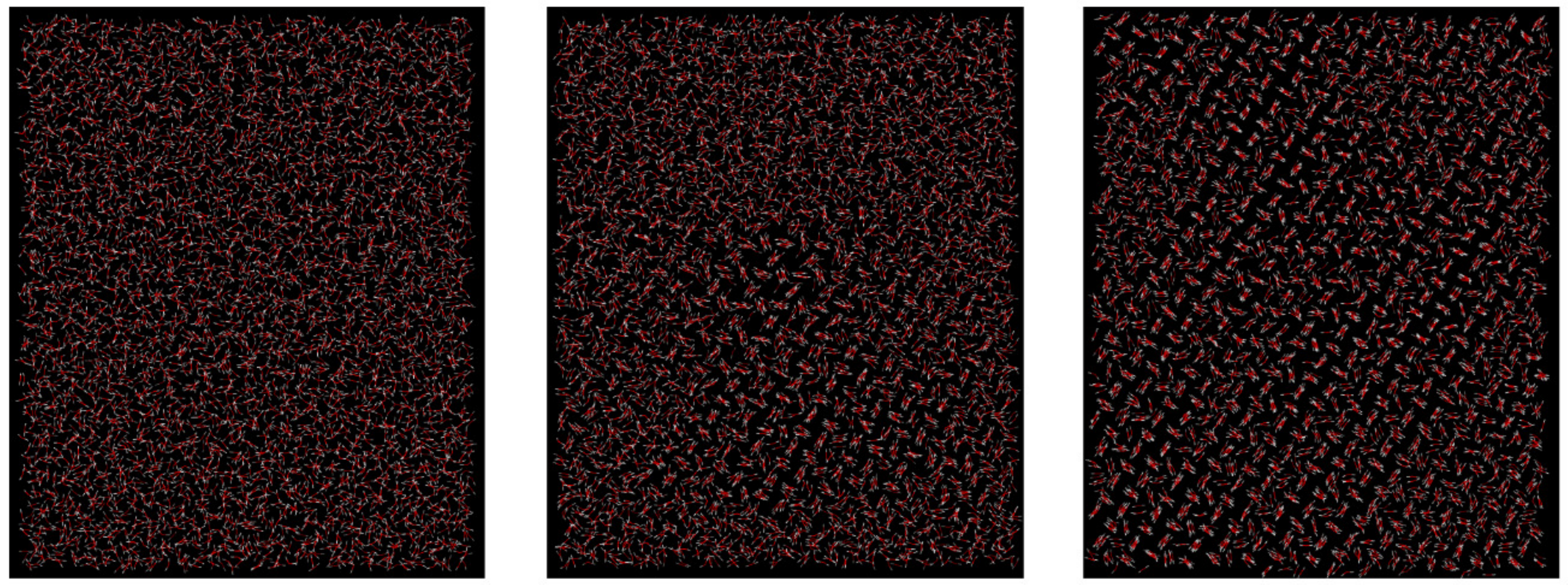}
\caption{Snapshots of a Molecular Dynamics simulation of TIP4P/ICE liquid water at 260 K under an electric field E=1.5 V/nm, 
showing homogeneous nucleation of ice Icf. The images correspond to times $t=0$ (left), $t=2.5$ ns (middle) and a later time 
when the simulation box has fully crystallized.}
\label{Snapshots}
\end{figure*}

However, Ice Ih is not the only phase that could nucleate when supercooling water at ambient pressure and under a large electric field. 
In fact, we have observed homogeneous nucleation 
of a ferroelectric cubic ice phase  (Icf) for homogeneous electric fields of $E$=1.5 V/nm using 
the TIP4P/ICE model, with the permanent polarisation vector of the growing ice Icf fully aligned with the direction of the field 
(at 260 K, nucleation spontaneously occurred in 2.5 ns, see Figure \ref{Snapshots}).
This phase was observed to nucleate with the field heterogeneously applied in Refs. \cite{Yan_2011,Yan_2012,Yan_2014}. 

Here, we have computed the melting temperature of ice Icf as a function of the applied field 
(red line in Fig. \ref{Fig_phase}). 
In our direct coexistence simulations we introduce a slab of ice Icf (ferroelectric and proton ordered) 
 in one side of the simulation box and liquid water in the other.
We determine the melting temperature as the largest temperature at which the solid does not melt. As we discuss in the Supporting Information, the melting line thus obtained corresponds to ice Ic at $E=0$, ice Icf for $E\ge0.3$ V/nm and to
partially polarized ice Ic for $0<E<0.3$ V/nm.
For this reason, we label the red melting curve in Fig. 2 as Ic/Icf.

The positive slope of the $T_m(E)$ line for ice Icf has been previously
reported in Ref. \cite{Yan_2014}. Here, we compare for the first time the $T_m(E)$ lines for both phases, 
showing a qualitatively different behaviour.
The difference between  both melting lines keeps increasing monotonically 
with the magnitude of the field, up to a difference of almost 150 K at $E=1.5$ V/nm. 
This result reveals that ice Icf becomes  thermodynamically more stable than ice Ih under large electric fields. 





Having established that ice Icf is thermodynamically more stable than ice Ih under a large constant electric field, it is also important 
to investigate how and whether an applied electric field  affects the nucleation  of ice Icf
from supercooled water. For that purpose, we  compute the nucleation rate of ice Icf
under an applied field of  0.3 V/nm,  corresponding to the same value of the field previously used in the study 
of the nucleation of ice Ih and also to the smallest value of the field
at which the difference between  ice Ih and ice Icf's  $T_m$  is clearly detectable, being the 
melting temperature for ice Icf at 0.3 V/nm 275$\pm$1 K (see Figure \ref{Fig_phase}). 

Following the same route as with ice Ih, we first compute the chemical potential difference between the supercooled 
liquid and ice Icf. The values of $\Delta \mu$ for ice Icf with the field (in red, see Figure \ref{Anuclea} (a)) 
are very similar to those of ice Ih without (in black) and with the field (in blue) down to a supercooling of about 30 K. 
Therefore, for a given supercooling, the thermodynamic driving force for nucleation of ice Ih and ice Icf is not strongly 
affected by the presence of the electric field. At higher supercoolings, the absolute value of $\Delta \mu$ for ice Icf 
under the field is 
higher than that of ice Ih without a field by more than 10\%, increasing the thermodynamic driving force for nucleation of 
the former with respect to the latter.

Next, we determine the size of the critical cluster $N_c$ as a function of supercooling for ice Icf under a field $E=0.3$ V/nm, 
shown on Figure \ref{Anuclea} (b) for comparison with ice Ih. Within the uncertainty of the simulation results,
clusters of ice Icf at 0.3 V/nm (red) are critical at nearly the same supercooling as those of ice Ih without a field (in black).
If we now compute ice Icf-water interfacial free-energy as a function of the supercooling (Figure \ref{Anuclea} (c) and 
Table \ref{table0}), we conclude that, for every supercooling,  $\gamma_{Ih}(E=0.0 $ V/nm) is quite similar to  
$\gamma_{Icf}(E=0.3$ V/nm), and both are lower than $\gamma_{Ih}(E=0.3$ V/nm). The Seeding values of $\gamma_{Icf}$ (points at $\Delta T>0$ in Fig. \ref{Anuclea}c) are 
consistent with those obtained by means of the Mold Integration method at coexistence (points at $\Delta T=0$ in Fig. \ref{Anuclea}c). This is a strong consistency test for our $\gamma$ calculations.

In order to calculate the attachment and nucleation rates of ice Icf under a field, shown on Table \ref{table0}, 
we have fixed the value of $\lambda\approx 4 \AA$ and used Equations \ref{fmas} and \ref{rate}. 
Finally, we compute the nucleation rate of ice Icf with the field by means of Eq. \ref{rate}, with results shown 
in Table \ref{table0} and Figure \ref{Bnuclea}.

We remind the reader that in the previous section we showed that a
0.3 V/nm electric field hinders ice nucleation via ice Ih.
However, if one considers Icf instead, the nucleation rate curve does not change within
the accuracy of our calculations (see black, Ih (E=0 V/nm), and red, Icf (E=0.3 V/nm), curves in Fig. \ref{Bnuclea}).
Thus, for a given supercooling, ice Icf with the field
nucleates at the same rate as Ih without the field.
In fact, both the nucleation driving force, $|\Delta \mu| (\Delta T)$, and the decelerating force, $\gamma (\Delta T)$,
do not change much from Ih(E=0 V/nm) to Icf (E=0.3 V/nm) (see black and red curves in Fig. \ref{Anuclea} a) and c)).
By contrast, when ice Ih is considered as the nucleating phase under the field, $\gamma$
raises significantly (see previous section). In that case we argued that the increase of $\gamma$ could be
due to a field-induced orientational misalignment
between water molecules belonging to the fluid (polarised) and those belonging to the ice Ih phase (non polarised). 
However, such misalignment is not present when ice Icf (polarised) is considered instead of ice Ih.

In summary, our calculations show that ice Ih without a field nucleates as fast as ice Icf under a 0.3 V/nm field
for a given supercooling. Then, the field would seemingly have a negligible effect on the speed with which 
ice crystals nucleate. However, because the melting temperature is higher with the field, 
if we consider absolute temperature rather than supercooling, a given nucleation
rate is reached at warmer temperatures. This leads us to the conclusion that the electric field
favours ice nucleation, in the form of ice Icf.     

We have shown in previous work \cite{Sanz_2013} that it is not accesible to observe homogenous nucleation of ice Ih in the absence of a field in brute force simulations. The spontaneous homogeneous nucleation of ice Icf observed in our simulations of water under electric fields of $E=$1.5 V/nm 
suggests that, at fields larger than 0.3 V/nm,
the nucleation rate $J$ of ice Icf must be significantly higher than that of ice Ih without field at the same supercooling. 
In the inset of
Figure \ref{Bnuclea}, it can be seen that, at very high supercoolings, 
$J$ of ice Icf at 0.3 V/nm is higher than that of ice Ih without the field and 
gets close to the threshold that would permit to observe spontaneous ice Icf nucleation in Molecular Dynamics simulations \cite{Sanz_2013}. Our simulations show that this threshold is clearly exceeded under a field of 1.5 V/nm at 260 K. 
In those conditions, the supercooling with respect to the melting temperature of ice Icf ($T_m=325$ K) 
is 65 K and the diffusion coefficient is still large enough to permit the nucleation of ice Icf clusters.
This justifies our observation of spontaneous homogeneous nucleation of ice Icf in liquid water at 260 K under a field
of 1.5 V/nm. 
In order to estimate the nucleation rate at 1.5 V/nm more accurately, the dependence of the ice Icf-water 
interfacial free-energy $\gamma$ and the chemical potential difference between ice Icf and water $\Delta\mu$ with the supercooling 
should be studied. 

We are aware of the existence of a polarized version of ice Ih, Ihf (in fact, this is denoted as ice XI\cite{JPCS_1986_47_00165_nolotengo}). We did not consider such structure in our study
because it must be less stable than ice Icf. The higher stability of a ferroelectric Ic phase with respect to a ferroelectric Ih phase is 
due to the fact that for ice Icf it is possible to obtain a value of $\langle M\rangle/(N \ \mu_{eff}) = 1$  (i.e full 
saturation) whereas for the Ihf the maximum value is $\langle M\rangle/(N \ \mu_{eff}) = 0.58$ due to the 
geometrical constraints of the lattice\cite{Aragones_2011a}.

\subsection{Effect of the electric field on the growth rate}

In order to predict the ability of ice formation one needs to know the rate of crystal growth as well as that of nucleation. With that purpose we compare in this section the speed of growth of ice Ih with no field with those of ice Ih and ice Icf under a 0.3V/nm field.
We perform  direct coexistence simulations of water-ice (exposing the secondary prismatic plane) at  different supercooling, and measure 
the  growth rate from the speed of the drop of the potential energy in the region in which it decays linearly during the crystallization process (see details in the Supplementary Information). 
\begin{figure}[!h]\centering
\includegraphics[clip,width=0.7\textwidth,angle=-0]{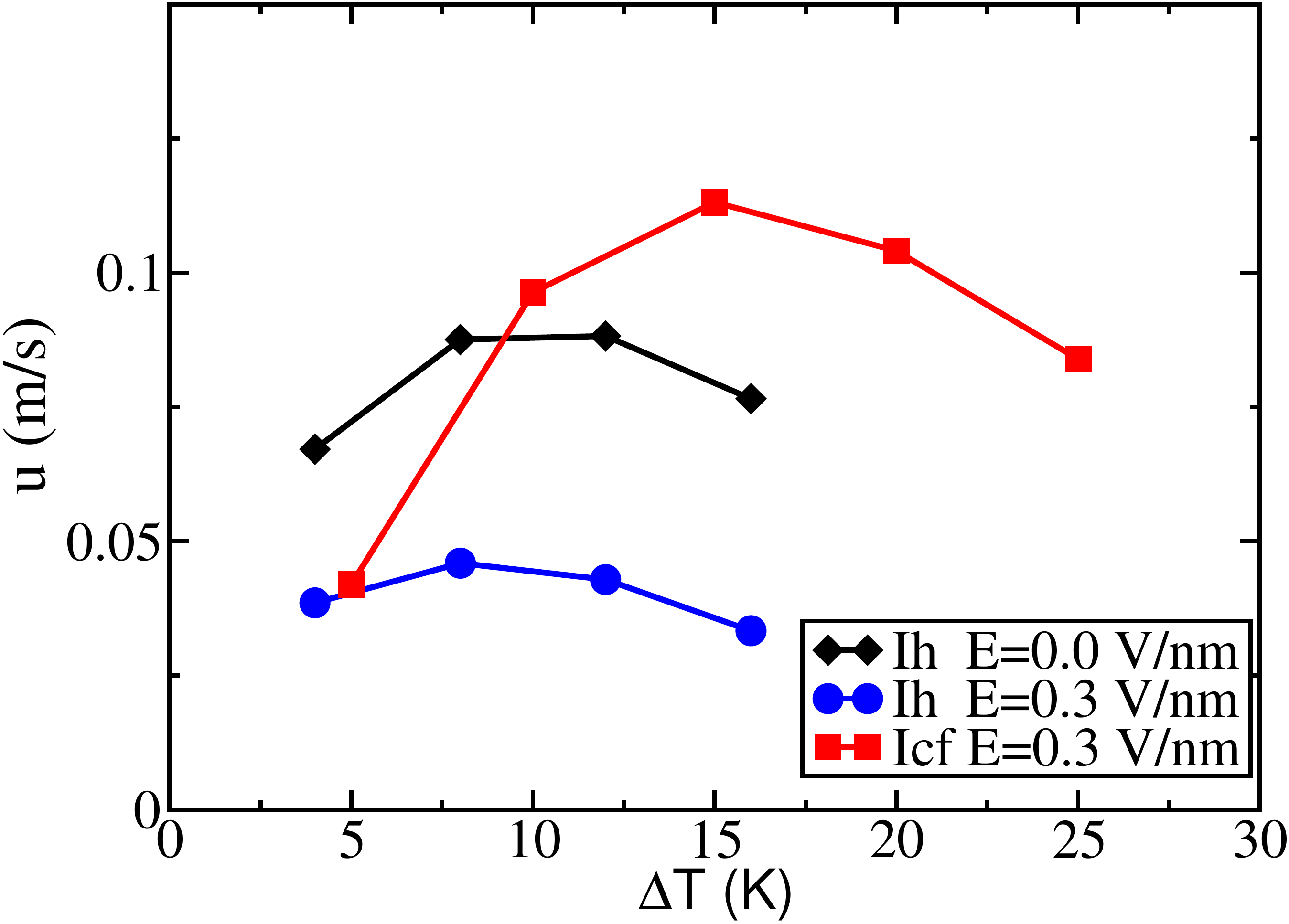}
\caption{Growth rates at 1 bar as as function of the supercooling of ice Ih at 0 V/nm (black symbols) and 0.3 V/nm (blue symbols) and ice Icf at 0.3 V/nm (red symbols). The melting points of each phase are: 
270 K (Ih, E=0 V/nm), 265.5 K (Ih, E=0.3 V/nm) and 275 K (Icf, E=0.3 V/nm).}
\label{Fig_growth}
\end{figure}

As shown in Fig. \ref{Fig_growth}, the growth rate $u$ of ice Ih without any applied field is a non-monotonic function of supercooling and shows a maximum 
at $\Delta T$ between 8 K and 12 K,  in  good agreement with previously published 
results by Espinosa at al\cite{Espinosa_2016} for TIP4P/ICE and by Rozmanov and Kusalik \cite{Rozmanov_2012a} for TIP4P/2005. 
Note that in this work we use a much simpler method to determine the growth rates than the one used by Rozmanov and Kusalik\cite{Rozmanov_2012a}.
When an electric field is applied, the growth of ice Ih is slowed down by a factor of two, and the supercooling that corresponds to the maximum in the growth rate is not strongly affected by the presence of the field. 
We hypothesise that the growth of ice Ih is hindered because the Ih solid is not polarised and the alignment of liquid molecules due to the field slows down the rate at which they can be incorporated into the growing crystal.
Simulation results for the TIP4P/2005 water model are presented in the Supporting information in agreement with the results for TIP4P/ICE.

When an electric field is applied, ice Icf grows faster than ice Ih. The growth rate $u$ is also a non-monotonic function of supercooling, with a maximum at around $\Delta T$=15 K. The effects of the electric field on the growth and nucleation rates are qualitatively similar. The field hinders the growth of ice Ih whereas it favours that of ice Icf. 
We hypothesise that the electric field favours the incorporation of molecules into the ice Icf phase because it aligns liquid molecules in the
direction they preferentially adopt in the solid, which is polarised in the direction of the field. 
Within the accuracy of our simulations, the growth rate of ice Ih without the field and that of ice Icf with an electric field of 0.3 V/nm are approximately equal in magnitude, but the maximum growth rate of ice Icf occurs at a deeper supercooling than for ice Ih. In terms of the absolute temperature, a strong electric field favours the growth of ice in the form of Icf. 
In any case, the quantitative effect of the electric field on the growth 
 rate is limited to a factor of 2, which is negligible compared with the effect on the nucleation rate, of many orders of magnitude.

\section{Discussion and Conclusions}

In this work we have numerically explored the effect of an electric field on homogeneous  ice nucleation, in order to assess the possibility of 
hindering homogeneous ice nucleation to support cryopreservation
(preventing the formation of both intracellular and extracellular ice crystals \cite{Fahy_1984}).
Even though it has been shown that the most important freezing mechanism for cryopreservation is heterogeneous ice nucleation\cite{morris,haymet}, 
in order to fully understand heterogeneous ice nucleation one has to first unravel the mechanism behind homogeneous ice nucleation.

Nucleation of ice Ih, the most stable polymorph (in the absence of an electric field) at ambient pressure,
is hindered when applying an electric field due to the increase of the ice-liquid interfacial free-energy. 
However, it is important to consider that when sufficiently large electric fields are applied the most stable polymorph is a polarised version of cubic ice, Icf (instead of ice Ih). 
When studying nucleation of ice Icf at a given electric field, we observe that its nucleation rate is comparable to the one of ice Ih at the same supercooling when no field is applied.
However, given that the melting temperature of ice Icf is higher than that of ice Ih, the field clearly favors ice nucleation, in the form of ice Icf. 

Even though, our results demonstrate that   reasonable electric fields  (smaller than the dielectric strength of water) are not relevant in the context of homogeneous ice nucleation at 1 bar, 
the electric field could still help cryopreservation. Switching on an external field in  
supercooled water could result in an instantaneous large increase of supercooling with respect 
to the ice Icf melting point, thus inducing  homogeneous nucleation of a large number of small ice Icf nuclei. 
In the presence of the electric field, ice Icf crystals grow faster than ice Ih, resulting in a solid of many small ice 
crystals, recently shown not to be detrimental for the cell's survival\cite{Huebinger_2016}. Further work is needed 
in order to understand the effect of an electric field on heterogeneous ice nucleation of water.

In Ref. \cite{Yan_2014}, spontaneous freezing of ice Icf was observed at 40 K supercooling for the six-site water model under 
strong electric fields, and it was argued that the size of the critical nucleus was determined by the degree of supercooling only 
and not by the magnitude of the field. Our simulations support that conclusion: the critical cluster size, the ice-water interfacial 
free-energy and the nucleation rate of ice Ih (when no field is applied) and those of ice Icf under a field of 0.3 V/nm are  
undistinguishable when plotted as a function of the supercooling. Larger values of the field should be studied in order to check 
if this similarity in the behaviour of ice Ih with no field and ice Icf with the field as a function of the supercooling still holds.

Our simulations
go in line with recent results on the effects of salt and pressure on homogeneous ice nucleation\cite{Espinosa_2017}. Pressure, 
salt and the electric field mainly affect ice nucleation by changing the ice-liquid interfacial free-energy.

It should be noted that the experimental value of the dielectric constant of ice Ih is larger than that of liquid water 
(the opposite occurring in the TIP4P/ICE model). Therefore a large electric field, such as those used in the present study, 
would stabilize the ice Ih phase with respect to water and increase its melting point. However, ice Icf would still be more stable because it has a very large  permanent polarisation. 

All the results presented in this work deal with DC electric fields. In order to unravel whether the nature of the electric field (whether constant or alternate) could affect the results, a few cases with both types of fields are compared in the Supporting information. Our preliminary results show that, in the limit of very high frequency, the melting temperature of ice Ih is the same as if there was no external field applied, whereas at very low frequency $T_m$ is similar to the case of a constant electric field. By changing the frequency, it is possible to shift the melting temperature continously between both limits.
Although the field magnitudes studied in this work are larger than the experimental dielectric strength of liquid 
water, our findings could pave the way for further studies on heterogeneous ice nucleation which could be relevant as 
alternative freezing routes in food industry or in 
cryopreservation of cells and organs.

\ack

This work was funded by grants FIS2013/43209-P, FIS2016-78117-P and FIS2016-78847-P of the MEC and the
UCM/Santander 910570 and PR26/16-10B-2. C.Valeriani 
acknowledges financial support from a Ramon y Cajal Fellowship. J. R. Espinosa acknowledges financial
support from the FPI grant BES-2014-067625. Calculations  were carried out in the
supercomputer facility  Magerit  from the Spanish Supercomputing Network (RES)
(project QCM-2015-1-0028). J. Ram\' irez acknowledges the computer resources and technical assistance 
provided by the Centro de Supercomputacion y Visualizacion de Madrid (CeSViMa).


\bibliographystyle{iopart-num}
\bibliography{./cluster,./ice,./bibliography,./bibliography_path_integral,./Additionalbib}

\end{document}